\documentclass[
aip,  jmp,
reprint,
amsmath,amssymb
]{revtex4-2}
\usepackage{amsthm}
\usepackage{amsfonts}
\usepackage{mathrsfs}

\theoremstyle{plain} 
\newtheorem{cor}{Corollary}

\newtheorem{thm}{Theorem}

\theoremstyle{definition}
\newtheorem{defn}{Definition}

\theoremstyle{remark}
  
\newtheorem*{case1.1}{\textbf{\small Case 1.1}}

\DeclareMathOperator{\ord}{ord}
\DeclareMathOperator{\res}{res}
\DeclareMathOperator{\reslog}{res log}
\DeclareMathOperator{\pd}{\partial}

\DeclareFontFamily{OT1}{pzc}{}
\DeclareFontShape{OT1}{pzc}{m}{it}{<-> s * [1.10] pzcmi7t}{}
\DeclareMathAlphabet{\mathpzc}{OT1}{pzc}{m}{it}

\DeclareMathSymbol{\R}{\mathalpha}{AMSb}{"52}
\DeclareMathSymbol{\C}{\mathalpha}{AMSb}{"43}
\newcommand{\mbb}[1]{\mathbb{#1}}
\newcommand{\Z}{\mbb{Z}}
\newcommand{\N}{\mbb{N}}

\newcommand{\set}[1]{\left\{#1\right\}}

\newcommand{\comment}[1]{}
\newcommand{\ra}{\rightarrow}

\newcommand{\bv}{\mathbf{v}}

\newcommand{\msr}{\mathscr{R}}

\newcommand{\mcp}{\mathcal{P}}

\newcommand{\mcr}{\mathcal{R}}
\newcommand{\mck}{\mathcal{K}}

\newcommand{\beq}{\begin{equation}}
\newcommand{\eeq}{\end{equation}}
\newcommand{\beqS}{\begin{equation*}}
\newcommand{\eeqS}{\end{equation*}}
\newcommand{\balign}{\begin{align}}
\newcommand{\ealign}{\end{align}}
\newcommand{\bsube}{\begin{subequations}}
\newcommand{\esube}{\end{subequations}}

\begin{document}


\title{Integrable classes of a family of  evolution equations} 



\author{J.C. Ndogmo}
\email[]{jean-claude.ndogmo@univen.ac.za}
\affiliation{Department of Mathematical and Computational Sciences\\
University of Venda\\
P/B X5050, Thohoyandou 0950, South Africa}


\date{\today}

\begin{abstract}
The problem of classification into symmetry integrable classes is solved for a  family of second order nonlinear evolution equations labeled by arbitrary functions. Four nonequivalent symmetry integrable classes are thus obtained and the results are transformed  into known integrable equations from the  literature. Recursion operators are also given for all the symmetry integrable classes found, some of which are the only ones known for some of the canonical classes.
\end{abstract}

\maketitle 

\section{Introduction}
\label{s:intro}

There are several concepts of integrability, but they all have fairly different meanings. A Hamiltonian system is called integrable if it has as many pairwise commuting integrals of motion as there are degrees of freedom in the system. On the other hand a given system of differential equations 
is said to be symmetry integrable if it has an infinite sequence of generalized symmetries of increasing orders. Another concept related to integrability is the painlev\'e Property. This refers to a system of differential equations whose solutions have poles as the only movable singularities. Although by their very nature all these concepts of integrability have different meanings, they all  express the strong potential that the underlying equation has for possessing a certain set of desirable properties which include linearizability. Indeed, linear equations are integrable in the sense of all the three concepts mentioned above, and they have all other usually desired properties for integrable systems. Such properties include the existence of a sufficiently large number of integrals of motion or conservation laws, generalized symmetries of increasing orders, Lax pairs or Darboux polynomials, the linearizability  by point or contact transformations, or the exact solvability by inverse scattering or differential substitution.\par

Despite the tremendously vast amount of research work accomplished within the framework of each of the various types of integrability, there are still however only a few logical connectives relating these various types, and in particular most often one cannot tell  if an equation integrable according to one type will necessarily  also be integrable according to one of the other types. \par

   Following the work of Shabat and co-workers, an algorithmic method has been found for identifying certain types of symmetry integrable equations \cite{shab1, shab2, shab3, mikh86}, based on the concept of so-called formal symmetry. This method yields in particular a classification of symmetry integrable equations from a given family of partial differential equations. Moreover, symmetry integrability has been found by many other authors  as one of the most suitable definition for integrability \cite{fokas87,mikh09,SW98,SW01,jirina,leznov}. Although the original method of Shabat and co-workers in the symmetry approach is more efficient for evolution equations in two independent variables, similar methods have been found for  non-evolutionary equations \cite{sokolov85, zhiber95, meshkov011}, as well as for {\sc ODE}s and more generally for equations on associative or non-associative algebras \cite{OS98, OW00, MS00, odess12}. Moreover, in parallel to these methods based on classical analysis, a purely algebraic approach often referred to as the symbolic method has also been developed  for certain types of equations including polynomial homogeneous equations, and applies to equations on both commutative and non-commutative algebras \cite{GD75, sh82, BCO81, Olv83}.

  We consider in this paper the  family of evolution equations of the form
\begin{equation}\label{e:main}
\frac{\pd u}{\pd t} = \Phi(u) \frac{\pd^2 u}{\pd x^2} + \Psi\left(u,\frac{\pd u}{\pd x}\right),
%
\end{equation}
where $u=u(t,x),$ and $\Phi$ and $\Psi$ are arbitrary  smooth functions of their arguments, with $\Phi\neq 0.$ We then perform a complete classification of this family of equations into symmetry integrable classes, under the most general point transformations mapping \eqref{e:main} into an equation of  the same form. \par

Although the classification into symmetry integrable equations of the most general family of second order evolution equations has been performed in  Ref. \onlinecite{svinolupov85}, such a classification does not provide all desirable information on the subclass \eqref{e:main}. Moreover, the classification performed in Ref. \onlinecite{svinolupov85} were done under contact transformations, and not point transformations as carried out in this paper. It turns out also that some the most commonly studied second-order  evolution equations are contained in the class \eqref{e:main}. In particular, this class contains many of the most important equations occurring  in physical and engineering  applications, including reaction-diffusion and Burgers equations and some of their variants (see Ref. \onlinecite{nimmo}). \par

In addition, not only we confirm the integrability of the nonequivalent classes we find by proving their equivalence to well-known integrable equations from the literature \cite{svinolupov85,sokolov_Arxiv1,SW01}, but we also determine, for the first time to the best of our knowledge, a recursion operator for some of the integrable classes found. As a result, recursion operators are now known for all integrable classes of \eqref{e:main}.



\section{Symmetry integrability}

  We summarize in this section some basic facts on symmetry integrability that will be needed in the sequel. Consider a scalar differential equation
\beq \label{e:meq}
\Delta \equiv \Delta (t, x, u_{(n)})= 0
\eeq
where $u=u(t,x)$ and $u_{(n)}$ denotes $u$ and all its partial derivatives with respect to $t$ and $x$ up to the order $n.$  Recall that $\Delta$ may be viewed as a differential function of $u$ in the sense of Ref. \onlinecite{olv86}. More generally, we denote by $\mck$ a differential field which will be assumed to be large enough to contain all the differential functions we wish to consider.  It is well known \cite{olv86} that a generalized vector field $\bv$ is a symmetry of \eqref{e:meq} if and only if the same holds for its evolutionary form $\bv_Q,$ where $Q$ is the characteristic. Thus the characteristic $Q$ is also called a symmetry of \eqref{e:meq}.

Let $D_t$ and $D_x$ denote the total $t$-derivative and $x$-derivative operators respectively, and $J=(n_1, n_2)$ the ordered multi-index given by $D_J = D_x^{n2} D_t^{n_1} ,$ where $ n_1, n_2 \in \Z^+.$ We set $\# J= n_1+ n_2.$  Let $u_J= D_{\!\!J} u,$ and for any differential function $Q$ of $u$ let $P^Q$ denote the differential operator given by
\beq \label{e:P^Q}
P^Q = \sum_{\#\, J\, \geq 0} D_J (Q) \frac{\pd}{\pd u_J}.
\eeq
In other words, $P^Q$ is the prolongation of infinite order of the corresponding evolutionary vector field $\bv_Q.$
\begin{defn} \mbox{}
\begin{enumerate}
\item[(a)] A function $Q\in \mck$ will be called a generalized symmetry of \eqref{e:meq} if
\[ P^Q (\Delta)_{|_{\Delta=0}}.\]
\item[(b)] Equation \eqref{e:meq} is called symmetry integrable if it possesses an infinite sequence of generalized symmetries of increasing orders.
\end{enumerate}
\end{defn}

Recall that the Fr\'echet derivative of a function $H\in \mck$ is the differential operator $D_H$ given by
\[
D_H =  \sum_{\#J\geq 0} \frac{\pd  H}{\pd u_J} D_J.
\]
Thus for any functions $Q, \Delta \in \mck$ one has $P^Q(\Delta) = D_\Delta (Q).$
We now restrict our attention to a special case of \eqref{e:meq} given by an evolution equation of the form
\begin{equation}\label{e:mee}
u_t= F(x, u, u_1, u_2, \dots, u_n)
\end{equation}
where we have set $u_i = \pd^i u/ \pd x^i$ for $i\geq 0$ and $u_0=u.$ We will often use the shorthand notation $F$ for the function $F(x, u, u_1, u_2, \dots, u_n).$  Restricting $\mck$ to the space of  functions $f= f(x, u, u_1, u_2, \dots, u_m),\; m\in  \Z^+,$ the total differential operator $D_x$ with respect to $x$ takes the form
\beq \label{e:Dx}
D_x = \frac{\pd}{\pd x} + \sum_{i=0}^{\infty}   u_{i+1} \frac{\pd }{\pd u_i}.
\eeq

Thus,   in particular on the space of functions defined on the solution set of \eqref{e:mee}, the operator $D_t$ is reduced to
\beq \label{e:Dt}
D_t= \frac{\pd}{\pd t} + \sum_{i=0}^{\infty} D_x^i (F) \frac{\pd }{\pd u_i}.
\eeq
The notation $D_x$ and $D_t$ will refer from now on to those given by \eqref{e:Dx} and \eqref{e:Dt}, unless otherwise stated. In particular, any symmetry $Q$ of \eqref{e:mee} satisfies $(D_t - D_F) Q=0.$\par

Denote by $\mck (D_x)$ the space of formal series $S$ of the form
\beq \label{e:formalS}
S= \sum_{j=-\infty}^m r_j D_x^j,\qquad r_j \in \mck,\quad r_m \neq 0, \qquad m \in \Z.
\eeq
The integer $m$ in \eqref{e:formalS} is called the order of $S$ and denoted $\ord (S).$ It is well known that $\mck (D_x)$  is a skewed field and in particular a non-commutative ring in which every non-zero element is invertible. Moreover, if $(r_m)^{1/m} \in \mck,$ then the $m$th root $S^{1/m}$ of $S$ exists in $\mck (D_x).$ The associative ring $\mck (D_x)$ is equipped with a Lie algebra structure for which the bracket is given by $[A,B]= A\circ B - B\circ A,$ where $A\circ B \equiv A B $ is the multiplication in $\mck (D_x).$

\begin{defn}\mbox{}
\begin{enumerate}
\item[(a)] A formal recursion operator of \eqref{e:mee} is a formal series $S\in \mck (D_x)$ satisfying
\beq \label{e:FROE}
P^F (S) - [D_F, S]=0.
\eeq
The set of all formal recursion operators of \eqref{e:mee} is denoted $\msr (F).$
\item[(b)] An approximate formal recursion operator of order $k$ of \eqref{e:mee} is an element of the set
\[ A_k = \set{S \in \mck (D_x) \colon  \ord(P^F (S) - [D_F, S]) \leq \ord (D_F) + \ord S - k}.\]
\end{enumerate}
\end{defn}
One readily sees that $\msr  (D_x) = A_\infty \subset \dots \subset A_3\subset A_2\subset A_1= \mck (D_x).$

\begin{defn}
Let $S$ be a formal series as in \eqref{e:formalS}.
\begin{enumerate}
\item[(a)]  The residue $\res (S)$ and the logarithmic residue $\reslog ( S )$ of $S$ are the functions given by
\[
\res(S) = r_{-1}, \qquad \quad \reslog (S) = r_{m-1} / r_m.
\]

\item[(b)] If $\ord (S)=m \neq 0,$ then the canonical densities $\rho_j,\; j=-1, 0,1, 2 \dots$ of \eqref{e:mee} associated with $S$ are the residues given by
\begin{align*}
\rho_{-1} &= \res S^{-1/m} = r_m^{-1/m},\quad \rho_0 = \reslog (S) = r_{m-1} /r_m \\
\rho_j &= \res S^{j/m},\qquad j\in \N.
\end{align*}
\end{enumerate}
\end{defn}

\begin{defn}
\begin{enumerate}
\item[(a)] Two  functions $f_1, f_2 \in \mck$ are said to be equivalent, denoted $f_1 \thicksim f_2,$ if $f_1-f_2 =D_x (h)$ for some $h\in \mck.$
\item[(b)] An element of the resulting quotient space $\bar{ \mck}= \mck / \!\thicksim $ is called a density.
\item[(c)] A non-zero element $\rho \in \bar{ \mck} $ is called a density of a local conservation law or a conserved density if $D_t (\rho) = D_x (\sigma)$ for some $ \sigma \in \mck.$ In this case $\sigma$ is called the flux and the pair $(\rho, \sigma)$ a local conservation law.
\end{enumerate}
\end{defn}

\begin{thm}  [\mbox{\rm [\onlinecite{mikh09}]}]
For each formal recursion operator $S \in \msr(F)$ with $\ord(S) \neq 0$ the associated canonical densities are densities of local conservation laws for $u_t=F,$ that is
\[  D_t (\rho_j) \in D_x (\mck),\qquad \text{ for all $j=-1,0,1,2, \dots$}. \]
\end{thm}
One of the immediate applications of this theorem is to readily check if an evolution equation has a formal recursion operator, and in particular a recursion operator in the ordinary sense of Ref. \onlinecite{olv86, SW01}. Recall that the order of a differential function $f$ of $u= u(t,x)$ is the highest order of the derivatives of $u$ appearing in the expression of $f.$

\begin{thm} [\mbox{\rm [\onlinecite{mikh09}]}]
If $u_t=F$ has a symmetry $H\in \mck$ of order $k,$ then $D_H \in A_k.$
\end{thm}

In particular, since the function $F$ as given by \eqref{e:mee} does not depend explicitly on $t,$ it is a symmetry for \eqref{e:mee} and hence $D_F \in A_n.$

\begin{cor} [\mbox{\rm [\onlinecite{mikh09}]}] \label{c:ARO} \mbox{}
\begin{enumerate}
\item[\rm{(a)}]  If $u_t=F$ has an infinite sequence of symmetries of increasing orders, then for any $\nu \in \Z$ there exists a formal recursion operator $S \in \mck (D_x)$ of order $\nu$ satisfying the formal recursion operator equation \eqref{e:FROE}.
\item[\rm{\rm{(b)}}] If $u_t=F$ has order $n \geq 2,$ then $u_t = F$ has an approximate formal recursion operator $S\in A_n.$

\item[\rm{(c)}]  If $u_t=F$ has order $n,$ then it has an approximate formal recursion operator of order $q> n$ if and only if the canonical densities satisfy
\beq \label{e:dsity-cdn}
P^F (\rho_j) = D_x (\sigma_j),\quad \text{ for $j= -1, 0, \dots, q-n-2,$}
\eeq
and for some fluxes $\sigma_j \in \mck.$
\end{enumerate}
\end{cor}

Given that every formal recursion operator is in particular an approximate recursion operator, it follows from the above corollary that if $u_t=F$ is symmetry integrable, then the canonical densities should satisfy \eqref{e:dsity-cdn} for all $q\in \Z.$

A recourse is usually made to approximate formal recursion operators in place of the formal recursion operators because in practice they are  easier to construct, and they include formal recursion operators as a special case. Moreover, Part (c) of Corollary \ref{c:ARO} asserts that the existence of an approximate formal recursion operator of order $ q >n$ is equivalent to the existence of $q-n$ local conservation laws given by \eqref{e:dsity-cdn}.


\section{Classification of symmetry integrable equations} \label{classif1}

   In order to perform a classification into symmetry integrable classes of the family of evolution equations of the form \eqref{e:mee},  one applies Conditions \eqref{e:dsity-cdn} and this yields the necessary conditions on the corresponding function $F$ for the equation to be symmetry integrable. However, Conditions \eqref{e:dsity-cdn} are so restrictive that the application of the first few of them usually yields all possible symmetry integrable equations \cite{mikh09, SW01}.   In the particular case of second order evolution equations,  and according to a result stated in Ref. \onlinecite{mikh09}, the first three canonical densities for the general second order evolution equation
\begin{subequations} \label{e:densitO2}
\begin{align}
u_t &=  F(x, u, u_1, u_2)  \label{e:gnl2ee}\\
\intertext{are given by}
\rho_{-1} &=  \left(  \frac{\pd F}{\pd u_2} \right)^{-1/2} \label{grho2m}\\
\rho_0 &=  \rho_{-1} \, \sigma_{-1} - \left(  \frac{\pd F}{\pd u_2} \right)^{-1}  \left(  \frac{\pd F}{\pd u_1} \right)                         \label{grho2z}\\
\rho_1 &=   \rho_{-1} \left(  \frac{\pd F}{\pd u}\right)  + \frac{\rho_0^2}{4 \rho_{-1}} + \frac{\rho_0 \, \sigma_{-1}}{2} - \frac{\rho_{-1}  \, \sigma_0}{ 2}.    \label{grho21}
\end{align}
\end{subequations}
We shall therefore apply \eqref{e:dsity-cdn} to the specific family of equations \eqref{e:main} using the expressions for the canonical densities provided in \eqref{e:densitO2}, to find conditions on the functions $\Phi$  and $\Psi$ for \eqref{e:main} to be symmetry integrable.\par

It follows from \eqref{e:densitO2} that for equation \eqref{e:main}, one has
\[
D_t \rho_{-1} = -\frac{\Phi_u (\Psi + \Phi u_2)}{2 \Phi^{3/2}}.
\]
On the order hand, one has $D_t \rho_{-1} \in D_x (\mck)$ if and only if $E(D_t \rho_{-1} )=0,$ where
\[
E= \frac{\delta }{\delta u} = \sum_{j=0}^\infty (-D_x)^j \frac{\pd }{\pd u_j}
\]
is the Euler operator acting on the space of functions $\mck.$ In order to determine  when $E(D_t \rho_{-1})=0,$ we are led to consider two cases.

\subsection*{Case 1: $\Phi_u \neq 0$}

In this case, It turns out that $D_t \rho_{-1}$ is a null Lagrangian, that is, $E(D_t \rho_{-1})=0,$ if and only if
\[
\Psi(u, u_1)= u_1 B + \frac{2 k_1 \Phi^{3/2} - u_1^2 \Phi'^2 + 2 u_1^2 \Phi'' }{2 \Phi'},
\]
where $B=B(u)$ is an arbitrary function. In order to evaluate $E(D_t \rho_j)$ for $j=0,1$ we would need to first evaluate the fluxes $\sigma_{i},\; -1\leq i <j $ appearing in \eqref{e:densitO2} in the expression of $\rho_j.$ Each $\sigma_i$ will be found by solving an equation of the form
\beq \label{e:detsigma}
D_t \rho_i - D_x \sigma_i(x, u, u_1, \dots, u_m) =0
\eeq
for the unknown function $\sigma_i=  \sigma_i(x, u, u_1, \dots, u_m),$ where $m-1$ is the order of $D_t \rho_i.$  This will usually be achieved by expanding \eqref{e:detsigma} as a polynomial in the derivatives $u_k$ of $u$ with $k>m.$ Thus solving \eqref{e:detsigma}  for $i=-1,$ yields
\[
\sigma_{-1} = -\frac{k_1 x}{2} -  \frac{u_1 f'}{2 \sqrt{f}} + \int_1^ u - \frac{B(z) f'(z)}{2 f(z) ^{3/2}} dz,
\]
where $k_1$ is an arbitrary constant of integration. With this value for $\sigma_{-1},$ the requirement that $D_t \rho_0$ must be a null Lagrangian gives rise to the condition
\beq \label{e:vder2dtrho0}
\begin{split}
0&= -B k_1 \Phi _u^3+8 u_2 \Phi ^{3/2} \Phi _u \left(-\Phi _u B_{uu}+B_u \Phi _{uu}\right)  \\
&\quad +4 \Phi ^{3/2} \big(\sqrt{\Phi } \left(-k_1
\Phi _u B_{uu}+k_1 B_u \Phi _{uu}\right)\\
&\quad +u_1^2 \left(-B_u \Phi _{uu}^2-\Phi _u^2 B_{uuu}+\Phi _u \left(B_{uu} \Phi _{uu}+B_u \Phi
_{uuu}\right)\right) \big).
\end{split}
\eeq
The requirement that the coefficient of $u_2$ in the above expression must identically vanish yields an expression for $B$ in terms of $\Phi,$ given by
\beq \label{e:B}
B= k_2 \Phi + k_3,\qquad \text{ for some arbitrary constants $k_2$ and $k_3.$}
\eeq
In terms of this new expression for $B,$ \eqref{e:vder2dtrho0} reduces to
\beq \label{e:vder2dtrho0V2}
-k_1 \Phi_u^3 B=0,
\eeq
and this leads to the consideration of  two other cases.

\subsubsection*{Case 1.1: $\Phi_u \neq 0,$  and $k_1=0.$}

Solving \eqref{e:detsigma} for $i=0$ and with the  value $k_1=0$ yields
\begin{equation*}
\begin{split}
\sigma_0 &= r_1-\frac{(f_1 k_2-k_3) (-k_3+k_2 \Phi )}{\sqrt{f_1} \sqrt{\Phi }}
+u_2 \left(\frac{\Phi _u}{2}-\frac{2
\Phi  \Phi _{uu}}{\Phi _u}\right) \\
&\quad +u_1 \left(\frac{\left(\sqrt{f_1} k_3+(-f_1 k_2+k_3) \sqrt{\Phi }+\sqrt{f_1}
k_2 \Phi \right) \Phi _u}{2 \sqrt{f_1} \Phi }-\frac{2 (k_3+k_2 \Phi ) \Phi _{uu}}{\Phi _u}\right)\\
&\quad +u_1^2 \left(-\frac{\Phi
_u^2}{4 \Phi }
+\frac{3 \Phi _{uu}}{2}-\frac{2 \Phi  \Phi _{uu}^2}{\Phi _u^2}\right),
\end{split}
\end{equation*}


where $r_1$ and $f_1$ are some constants of integration. Now that $\sigma_0$ is known, we can evaluate $\rho_1$ and then $E(D_t \rho_1).$ The identical vanishing of the coefficient of $u_4$ in the expression of $E(D_t \rho_1)$ shows that $E(D_t \rho_1)=0$ if and only if
\[
\Phi= p_2(3 u - 8 p_1)^{8/3},
\]
 where $p_1$ and  $p_2$ are some arbitrary constants. For the expressions thus obtained for $\Psi$ and $\Phi,$ the corresponding reduced equation \eqref{e:main} takes the form
 \beq \label{e:candidat1}
 u_t=  p_2 (-8 p_1+3
u)^{8/3} u_2 + p_2 (-8 p_1+3 u)^{5/3} u_1^2     -u_1\left(-k_3-k_2 p_2 (-8 p_1+3 u)^{8/3}\right).
 \eeq
In the sequel, unless otherwise stated the renaming of new variables to former ones under point transformations will be assumed in the transformed equation. Performing the change of dependent variable $u= \left[ 8 p_1 + (w/p_2)^{3/8}  \right]/3$  transforms  \eqref{e:candidat1} into
 \beq \label{e:intgrab1}
u_t = (k_3 + k_2 u) u_1 - \frac{u_1^2}{2} + u u_2.
\eeq

However, the later transformation is equivalent under point transformation
\beq \label{e:typeq1}
u_t = u^2 u_2.
\eeq
Indeed, for $k_2 \neq 0$ the corresponding transformation is given by
\[
x= -k_3 t + \frac{1}{k_2}\log(k_2\, z),\qquad u = \frac{w(t,z)^2}{k_2^2\,  z^2},
\]
while for $k_2=0$ it is given by
\[
x= -k_3 t + z,\qquad u= w(t,z)^2.
\]

\subsubsection*{Case 1.2: $\Phi_u \neq 0,$ and $k_1 \neq 0.$}
In this case, $B=0$ must hold. Without any further condition on $\Phi,$ it turns out that $D_t \rho_{0}$ is a null Lagrangian, and the corresponding flux takes the form
\beq \label{e:sigma1.2}
\begin{split}
\sigma_0 &= b_1 + \frac{k_1^2 x^2}{8} + \frac{k_1 x \Phi_u u_1}{4 \sqrt{\Phi}} - \frac{2 k_1 \Phi^{3/2} \Phi_{uu} }{\Phi_u^2} \\
&\quad + u_1^2 \left( -\frac{\Phi_u^2}{4 \Phi}  + \frac{3 \Phi_{uu}}{2} - \frac{2 \Phi \Phi_{uu}^2}{\Phi_u^2}  \right) + u_2\left( \frac{\Phi_u}{2} - \frac{2 \Phi \Phi_{uu}}{\Phi_u}   \right),
\end{split}
\eeq
 where $b_1$ is an arbitrary constant. With this expression for $\sigma_0$ and other relevant parameters computed thus far, we can compute $\rho_1$ in \eqref{e:densitO2} and then $E(D_t \rho_1).$ The identical vanishing of the coefficient of $u_4$ in the equation $E(D_t \rho_1)= 0 $ shows that  $D_t \rho_1$ is a null Lagrangian if and only if
\[
\Phi= p_2 (3 u- 8 p_1)^{8/3},
\]
as in the preceding case,  where $p_1$ and  $p_2$ are some arbitrary constants. The corresponding equation \eqref{e:main} is reduced to
\beq \label{e:eqcas1.2S}
 u_t=   \frac{k_1}{8} p_2^{1/2} (-8 p_1 +3 u)^{7/3} + p_2 (-8 p_1 +3 u)^{5/3} u_1^2 + p_2 (-8 p_1 +3 u)^{8/3} u_2.
\eeq

Performing the change of dependent variables $t= 2y/k_1,\; x= (2z/k_1)^{1/2}$ and  $u= \left[ 8 p_1 + (w/p_2^{1/2})^{3/4}  \right]/3$  transforms \eqref{e:eqcas1.2S} to

\beq \label{e:typeq2}
 u_t=    u^2(1 + u_2).
\eeq

\subsection*{Case 2: $\Phi_u=0.$}
Given that, $\Phi$ is a nonzero constant, by the scaling transformation $t \ra t/\Phi$ we may assume without loss of generality that $\Phi=1.$ In this case, $\rho_{-1}=1,$ and hence $\sigma_{-1}$ is a function of $t.$ In the specific case of \eqref{e:main}, since  the densities and fluxes cannot depend explicitly on $t$ and the fluxes are determined only up to an arbitrary constant, we may assume $\sigma_{-1}=0.$  This yields $\rho_0= -\Psi_{u1}.$ In the expansion of $E(D_t \rho_0)$ as a polynomial in the derivatives of $u,$ the vanishing of the coefficient of $u_4$ shows that $\Psi$ must be quadratic in $u_1,$ that is,
\beq \label{e:psi-case2}
\Psi= \alpha + \beta u_1 +\gamma u_1^2,
\eeq
for some functions $\alpha,$ $\beta,$ and $\gamma$ of $u.$ From this point we have to consider the following  subcases.

\subsubsection*{Case 2.1: $\Phi_u=0$ and $\beta_{u} \neq 0. $}
Substituting the expression of $\Psi$ from \eqref{e:psi-case2} in the expression for $D_t\, \rho_0$ shows that $E(D_t\, \rho_0)=0$ if and only if
\beq \label{e:beta cas2.1}
\Psi= \frac{d_1}{\beta_u} + \beta u_1 + \frac{\beta_{uu}}{\beta_u} u_1^2
\eeq
for some arbitrary constant $d_1.$ The corresponding expression for $\sigma_0$ takes the form
\beqS
\sigma_0= -\frac{\left(\beta^2+2 d_1 x\right) \beta_u^2+2 \beta_u^3 u_1+4 \beta_{uu} \left(d_1+u_1^2 \beta_{uu}\right)+4 \beta_u \beta_{uu} \left(\beta u_1+u_2\right)}{2
\beta_u^2}.
\eeqS
With this expression for $\sigma_0,$ we can then compute $\rho_1,$ and it turns out that without any further restrictions on $\Psi,$ $D_t\, \rho_1$ is a null Lagrangian. The corresponding reduced equation in this case takes the form
\beq \label{e:eqcas2.1}
u_t = u_2 + u_1^2 \frac{\beta_{uu}}{\beta_u} + u_1 \beta + \frac{d_1}{\beta_u}.
\eeq
Moreover, the change of dependent variable $w= \beta(u)$ maps the reduced polynomial equation
\beq \label{e:eqcas2.1V2}
u_t = u_2 + u u_1+ d_1
\eeq
precisely to \eqref{e:eqcas2.1}. In addition, the  transformation
\[
t= 1+ 1/y,\qquad x= -\frac{d_1}{2 y^2} + \frac{z}{y} i,\qquad u = - i \left(  \frac{d_1}{y}i + z + y w(y,z) \right),
\]
where $i$ is the imaginary number,  maps  \eqref{e:eqcas2.1V2} to its homogeneous version
\beq \label{e:burgerseq}
u_t = u_2 + u u_1.
\eeq

\subsubsection*{Case 2.2: $\Phi_u=0$ and $\beta_{u}= 0. $}
 In this case denoting by $\beta_0$ the constant function $\beta$, we have $\rho_0 = -\beta_0 - 2 u_1 \gamma,$  and this yields
\beq
\sigma_0 = p_3 - 2 \alpha \gamma -2 \gamma ^2 u_1^2 - 2 \gamma (\beta_0 u_1 + u_2).
\eeq
For this value of $\sigma_0,$ we can compute $\rho_1.$ It turns out that $D_t \rho_1$ is a null Lagrangian for $\alpha=0,$ and the corresponding equations takes the form
\beq \label{e:eqcas2.2.1S}
u_t =  u_2 + \beta_0 u_1 + \gamma u_1^2 .
\eeq
The change of variables
\[
t= z,\qquad  x= \beta_0 - \beta_0 z - y,\qquad  u= \int_1^w e^{\int_1^{v_1} \gamma (v_2) d v_2} d v_1,
\]

after reverting back to the original variables, transforms the (linear) heat  equation

\beq \label{e:heateq}
u_t =  u_2
\eeq
precisely to \eqref{e:eqcas2.2.1S}, showing indeed that \eqref{e:eqcas2.2.1S} is linearizable.  On the other hand, if  $\alpha \neq 0,$  then $D_t \rho_1$ is a null Lagrangian if and only if $\gamma = - \alpha_u/ \alpha,$  and the corresponding equation takes the form
\beq \label{e:eqcas2.2.2}
u_t =  u_2 + \beta_0 u_1 + \alpha - \frac{\alpha_u}{\alpha} u_1^2.
\eeq
On the other hand, the change of variables $z= x+ \beta_0 t$ and $w= \int^u \frac{1}{\alpha(s)}  d s,$ with $w=w(y,z),$ maps the  heat equation \eqref{e:heateq} precisely to \eqref{e:eqcas2.2.2}, after a renaming of variables in the transformed equation.


\subsection{Integrable equations} \label{ss:integrableq} We now proceed to the determination of integrable equations from among the four candidates obtained in our preliminary classification of integrable equations, namely the equations  \eqref{e:typeq1}, \eqref{e:typeq2}, \eqref{e:burgerseq}, and \eqref{e:heateq}. Indeed, these four equations result only from an application of the necessary conditions of integrability.\par

By essence, and as a basic fact of integrability, linear equations are integrable and hence this holds true in particular for the heat equation \eqref{e:heateq}. Consider now  the general second order evolution equation \eqref{e:gnl2ee}, namely
\begin{equation*}
u_t = F(x, u, u_1, u_2)
\end{equation*}
where $F$ is an arbitrary smooth function of its arguments. This is indeed a more general family of equations which contains the family of equation \eqref{e:main} whose symmetry classification is discussed in this paper. According to a result reported in Ref. \onlinecite{sokolov_Arxiv1}, the exhaustive list of integrable equations of the form \eqref{e:gnl2ee} which are not linearizable by \emph{contact transformations} had been found in Ref. \onlinecite{svinolupov85}, and it is given by
\begin{subequations}\label{e:cl2nlee}
\begin{align}
 u_t &= u_2 + 2 u\, u1 + h(x) \label{e:cl2nlee1} \\
 u_t &= D_x \big(u_1\, u^{-2} +  \alpha x\, u + \beta u  \big) \label{e:cl2nlee2} \\
 u_t &= D_x \big(u_1\, u^{-2} - 2 x   \big), \label{e:cl2nlee3}
\end{align}
\end{subequations}
where $h$ is an arbitrary smooth function of $x$ while $\alpha$ and $\beta$ are arbitrary constants. It should be mentioned that for evolutions equations of the form \eqref{e:gnl2ee} in which the function $F$ does not depend explicitly on the variable $t,$ the contact transformations just referred to can be taken in the form
\begin{subequations}
\begin{align}
z = t,\qquad y &= \phi(x, u, u_1),\qquad w = \psi(x, u, u_1) \\
\intertext{ in which the functions $\phi$ and $\psi$ are constrained by the contact condition}
D_x (\phi) \frac{\pd \psi}{\pd u_1} &= D_x(\psi) \frac{\pd \phi}{\pd u_1}.
\end{align}
\end{subequations}

 It is much easier at this point to make use of \eqref{e:cl2nlee} to conclude on the integrability of the candidate equations, although we will also use other arguments to prove integrability.\par

First letting $h=0$ in \eqref{e:cl2nlee1} and then applying the change of dependent variable $w = 2 u$ transforms the resulting equation with $h=0$ into Burgers equation \eqref{e:burgerseq}, and thus confirms its well-known integrability.
On the other hand, the point transformation
\[
 t= z,\qquad x = 2y,\qquad u = 2/w
 \]
maps equation \eqref{e:typeq2} precisely to \eqref{e:cl2nlee2}, showing its integrability according to Ref. \onlinecite{svinolupov85}. Finally, the mere change of dependent variable $u= 1/w$ maps \eqref{e:typeq1} precisely to \eqref{e:cl2nlee2} with $\alpha=\beta=0.$ This proves also the integrability of \eqref{e:typeq1} according to Ref. \onlinecite{svinolupov85}. We have thus shown that our three candidate nonlinear equations are all integrable.\par

\begin{thm}\label{t:summary} \mbox{}
\begin{enumerate}
\itemsep=1.5mm
\item[{\rm (a)}] Any quasilinear evolution equation of the form \eqref{e:main} is symmetry integrable if and only if it is equivalent under point transformations to one of the following equations
\begin{subequations} \label{e:classif}
\begin{align}
u_t &= u_2      \label{e:heateq2}  \\
u_t &= u^2 u_2      \label{e:typeq1S}  \\
u_t &= u^2( 1+u_2)      \label{e:typeq2S}  \\
u_t &= u_2 + u u_1       \label{e:burgersS}.
\end{align}
\end{subequations}

\item[{\rm  (b)}] The four equations in \eqref{e:classif} are pairwise nonequivalent under both point transformations and contact transformations.
\end{enumerate}
\end{thm}
\begin{proof}
The proof that any symmetry integrable equation of the form \eqref{e:main} must be equivalent under point transformations to one of the equations \eqref{e:classif} is proved in the first part of Section \ref{classif1}, while the fact that all equations \eqref{e:classif} are symmetry integrable has been proved above in the present Subsection \ref{ss:integrableq}. The fact that  all equations in \eqref{e:classif} are nonequivalent under point transformations can be easily verified by the fact that their symmetry algebras all have distinct dimensions. We have also proved that all  the equations in \eqref{e:classif} are nonequivalent under contact transformations, by mapping each of them to nonequivalent symmetry integrable equations obtained in Ref. \onlinecite{svinolupov85} under contact transformations.
\end{proof}

\section{Recursion operators}

An alternative and more direct way to establish  integrability  is to find a recursion operator  for the concerned equation. Let $\Delta[u] =0$ be a system of differential equations and denote as usual by $\bv_Q$ a generalized symmetry with characteristic $Q$ of this system.  Recall that a pseudo-differential operator $\mcr$ acting on a space of differential functions is called a recursion operator for $\Delta[u]=0$ if $\bv_{\tilde{Q}}$ is also a generalized symmetry for this system, where $\tilde{Q}= \mcr\, Q,$ for some initial characteristic function $Q.$ Indeed, if $\mcr$ is a recursion operator,  $\bv_Q$ is any generalized symmetry for $\Delta$ and $\tilde{Q}= \mcr\, Q,$ then $\bv_{\tilde{Q}}$ is not necessarily also a generalized symmetry for $\Delta.$ However, for a given recursion operator there is a sequence of generalized symmetries $\bv_{Q_j}$ of $\Delta$ given by $Q_{j+1} = \mcr\, Q_j$ for $j\in \N=\set{1,2, 3, \dots}.$ The first characteristic $Q_1$ in this recursively defined sequence of characteristics is referred to as the root of $\mcr.$

 The best known and most documented method for finding these operators is probably the method given in Ref. \onlinecite{olv86}. However, this method involves a lot of guess work regarding in particular the highest and the lowest orders of derivatives of terms in the operator, as well as the number and types of variables that it should depend on. All this makes the determination of recursion operators a particularly tricky task. Another method for finding recursion operators appears in Ref. \onlinecite{SW01}, but as this second method is not sufficiently documented we shall make use of the method of Ref. \onlinecite{olv86}. \par

First of all, let us recall that recursion operators  for the heat equation \eqref{e:heateq2} and for Burgers equation \eqref{e:burgersS} are well known \cite{olv86,SW01}. For instance $D_x$ and $t D_x + \frac{1}{2}x$ are two recursion operators for \eqref{e:heateq2}. More generally, criteria for identifying recursion operators for linear equations are much more relaxed, and they are easier to find. A recursion operator  for \eqref{e:burgersS} has expression $D_x + \frac{1}{2}u - \frac{1}{2} u_1 D_x^{-1}.$ In order to provide an exhaustive list of recursion operators for the  classes of integrable equations \eqref{e:classif}, we are only left to do this for equations \eqref{e:typeq1} and  \eqref{e:typeq2}. To our knowledge recursion operators for these two equations are not available in the current literature.\par

By a result of  Olver\cite[Theorem 5.30]{olv86}, for a pseudo-differential operator $\mcr$ to be a recursion operator of an evolution equation $\Delta[u]=0,$ it suffices that it commutes on solutions with the Fr\'echet derivative $D_\Delta$ of $\Delta,$ that is $\mcr \cdot \Delta= \Delta \cdot \mcr$ for all solutions $u$ to $\Delta.$\par

After some calculations, it turns out that
\begin{equation}\label{e:recurtyp1}
\mcr = u D_x+ u^2 u_2 D_x^{-1} \cdot \left( \frac{1}{u^2} \right)
\end{equation}
is a recursion operator of root $u^2 u_2$ for \eqref{e:typeq1}. Indeed, the Fr\'echet derivative
\[
D_\Delta = -D_t + u^2 D_x^2 + 2\, u\, u_2
\]
of \eqref{e:typeq1} commutes with $\mcr.$ To verify this we set
\begin{equation}
\mcr = \mcr_1 + \mcr_2,\quad \text{ where }\quad \mcr_1= u D_x,\quad \text{ and }\quad  \mcr_2 = u^2 u_2 D_x^{-1}\cdot \left( \frac{1}{u^2} \right).
\end{equation}
A straightforward  calculation shows that
\begin{equation}
D_\Delta  \cdot \mcr_1  -\mcr_1 \cdot D_\Delta = - 2 u (u_1 u_2 + u u_3). \label{e:fr1f}
\end{equation}
On the other hand similar calculations show that
\begin{subequations}\label{e:comt1eq}
\begin{align}
\begin{split} \label{e:fr2}
D_\Delta \cdot \mcr_2 &=  u \bigg(4 u^2 \left(u_2^2+u_1 u_3\right) D_x^{-1} \cdot \left( \frac{1}{u^2} \right)\\
&+ 2 u_2 \left(2 u_1-u^2 u_2 D_x^{-1} \cdot \left( \frac{1}{u^2} \right)\right)\\
&+u^3 \left(u_2 D_x \cdot \left(\frac{1}{u^2}\right)+u_4 D_x^{-1} \cdot \left( \frac{1}{u^2} \right) \right) \\
& -u \bigg(\big(-2 u_1^2 u_2+u_2 \left(2 u_1^2+2 u u_2\right)\\
&+4 u u_1 u_3+u^2 u_4\big) D_x^{-1} \cdot \left( \frac{1}{u^2} \right) \\
&-2 u_3 +u_2 D_t D_x^{-1} \cdot \left( \frac{1}{u^2} \right) \bigg)\bigg)=0 \\
\end{split}
\intertext{ and }
\mcr_2 \cdot D_\Delta &= u^2 u_2 \left(D_x+D_x^{-1} \cdot \left(\frac{2 u_2}{u}\right)+D_x^{-1} \cdot \left(-\frac{1}{u^2} D_t \right)\right). \label{e:r2f}
\end{align}
\end{subequations}

It thus follows from \eqref{e:fr2} and \eqref{e:r2f} that
\begin{equation} \label{e:fr2f}
\begin{split}
D_\Delta  \cdot \mcr_2  -\mcr_2 \cdot D_\Delta  &= u \bigg(4 u_1u_2-u u_2D_x+u^3 u_2D_x \cdot \left(\frac{1}{u^2}\right) \\
 &\quad -u \big(-2 u_3+u_2D_t D_x^{-1} \cdot \left(\frac{1}{u^2}\right)\\
&\quad +u_2
D_x^{-1} \cdot \left(\frac{2 u_2}{u}\right) +u_2D_x^{-1} \cdot \left(-\frac{1}{u^2} D_t \right)\big)\bigg).
\end{split}
\end{equation}
Combining \eqref{e:fr1f} and \eqref{e:fr2f} gives
\begin{equation}\label{e:fr2fv2}
\begin{split}
D_\Delta  \cdot \mcr  -\mcr \cdot D_\Delta  &= u u_2 \bigg(2 u_1-u D_x+u^3 D_x \cdot \left(\frac{1}{u^2}\right)\\
&\quad -u  D_x^{-1} \cdot \left( D_t \cdot \left( \frac{1}{u^2}\right)   + \frac{2 u_2}{u}  - \frac{1}{u^2} D_t    \right) \bigg).
\end{split}
\end{equation}
But the left hand side of \eqref{e:fr2fv2} vanishes since
\[
u^3 D_x \cdot \left(\frac{1}{u^2} \right) = -2 u_1+u D_x,\qquad \text{ and }  D_t \cdot \left( \frac{1}{u^2}\right)  = - \frac{2 u_2}{u}  + \frac{1}{u^2} D_t.
\]
This completes the proof that the expression $\mcr$ in  \eqref{e:recurtyp1} is a recursion operator for \eqref{e:typeq1}, and in particular that \eqref{e:typeq1} is integrable. Indeed, if we set $Q_1= u^2 u_2,$ then $Q_2= u^2 (3 u_1 u_2 + u u_3)$  is of higher order than $Q_1$ and more generally it is easy to verify that the sequence of characteristics $Q_{j+1}:= \mcr \, Q_j$ is of increasing orders.\par

In order to be guided in the search of recursion operators, it is always helpful to have at hand the first few values of characteristics of the concerned equation. In the case of Equation \eqref{e:typeq2},  characteristics $C_j$ of order not exceeding the fourth are given by
\begin{subequations}\label{e:char4type2}
\begin{align}
\begin{split}
 C_1 &= \frac{1}{2} u^2 \bigg( 2 u^2 u_4 + 2 u (3 + 7 u_2 + 4 u_2^2 + 6 u_1 u_3 + 2 u_3 x) \\
& + (1+ u_2) (14 u_1^2+ 12 u_1 x+ 3 x^2) \bigg)
\end{split}
\end{align}
\vspace{-10mm}
\begin{alignat}{2}
 \intertext{\vspace{-2mm}and} \notag
   C_2&= u_1,             &\quad  C_3 &= \frac{1}{2} u^2\left(  6 u_1 (1+ u_2) + 2 u u_3 + 3x(1+ u_2)\right) \\
   C_4&= u^2 (1+ u_2),    &\quad  C_5 &=  -2 u - 2 t u^2 (1+ u2) + u_1 x.
\end{alignat}

\end{subequations}
Partly guided  in some ways by these first few low order characteristics of generalized symmetries of \eqref{e:typeq2}, it is found after some calculations that \eqref{e:typeq2} has a recursion operator $\mcp$ with expression
\begin{equation} \label{e:recurtyp2}
\mcp =  u D_x  + \left( \frac{ x}{2}   \right) +    u^2 (1+ u_2) D_x^{-1} \cdot \left( \frac{1}{u^2} \right)
\end{equation}
and root $u^2 (1+ u_2).$ This fact is also established by proving that $\mcp$ commutes with the Fr\'echet derivative
\[
D_\Delta = D_t - u^2 D_x^2 - 2 u (1+ u2)
\]
of \eqref{e:typeq2}. The proof of this commutativity is  similar to the one given above to establish  that the expression $\mcr$ in \eqref{e:recurtyp1} is a recursion operator for  \eqref{e:typeq1}, and the details are  omitted. Let us mention also that the sequence of characteristics recursively defined by $\mcp$ is of increasing order, giving another proof of the fact that \eqref{e:typeq2} is symmetry integrable. For instance, while the root $C_4= u^2(1+ u_2)$ is of order $2,$ $\mcp\, C_4=  C_3$ is of order $3.$\par




\section{Concluding Remarks}
\label{s:conclusion}

Our classification results  show that the particular reduced class \eqref{e:main} contains all four subclasses and essentially  all  symmetry integrable equations classified under contact transformations in Ref.  \onlinecite{svinolupov85} for the most general class \eqref{e:gnl2ee}. This suggests that the classification of Ref. \onlinecite{svinolupov85} would certainly have yielded the same results under the much simpler and much common point transformation. On the other hand, our results also raise the problem of existence and determination of the smallest subclass of \eqref{e:gnl2ee} containing all integrable equations from \eqref{e:gnl2ee}, that is, all equations \eqref{e:cl2nlee} and \eqref{e:heateq2}.\par

This work has also revealed that despite the crucial importance of recursion operators in the study of integrable systems, their determination remains a very tricky task and considerable research work is still needed to understand some of their most basic properties.  Recursion operators $\mcr$ for a scalar evolution equation in two independent variable $t$ and $x$ as in the present case are to be sought in the form $\mcr= \sum_{k= m_1}^{m_2} F_k[u] D_x^{r_k},$ where $m_1,\, m_2,$ and $r_k$ are integers, some or all of which may assume negative values. However, the determination of crucial  profiling parameters of $\mcr$ such as $m_1,$ $m_2,$ and even the type and number of arguments for each function $F_k$ remains largely  the result of guess work \cite{olv86}, leaving the determination of $\mcr$ to remain really challenging.\par

\section*{Data availability statement}
\label{s:data}
Data sharing is not applicable to this article as no new data were created or analyzed in this study.

\section*{Acknowledgements}
\label{s:acknowledge}
Funding: This work was supported by the NRF Incentive Funding for Rated Researchers grant [Grant Number 97822];
the University of Venda [Grant Number I538].

\bibliographystyle{model1-num-names}


%
%

%



\end{document}